\documentclass[conference, 10pt]{IEEEtran}

\IEEEoverridecommandlockouts

\usepackage{color}
\usepackage{bm}

\usepackage{color}
\usepackage{theorem}
\usepackage{times,amsmath,epsfig}
\usepackage{amssymb}
\usepackage{subfigure}
\usepackage{cite}
\usepackage{url}

\usepackage{algorithmic}
\usepackage{algorithm}

\usepackage{tikz}
\usetikzlibrary{shapes,arrows}
\input{mysymbol.sty}

\newcommand{\QED}{\hfill\ensuremath{\blacksquare}}

\title{Conversational Turn-taking as a Stochastic Process on Networks}
\author{Lisa O’Bryan\textsuperscript{1}, Santiago Segarra\textsuperscript{2}, Jensine Paoletti\textsuperscript{1}, Stephanie Zajac\textsuperscript{1}, Margaret E. Beier\textsuperscript{1}, \\ Ashutosh Sabharwal\textsuperscript{2}, Matthew Wettergreen\textsuperscript{3}, and Eduardo Salas\textsuperscript{1}

\thanks{\textsuperscript{1}Department of Psychological Sciences, Rice University, Houston, TX, USA.
\textsuperscript{2}Department of Electrical and Computer Engineering, Rice University, Houston, TX, USA.
\textsuperscript{3}Department of Bioengineering, Rice University, Houston, TX, USA.
Subsets of the team data used in this manuscript have been previously published as part of dissertations by Stephanie Zajac, Department of Psychological Sciences, Rice University and Jian Cao, Department of Electrical and Computer Engineering, Rice University. Funding for this project was provided by a Microsoft Productivity Research Grant, the National Science Foundation (Award Number: 1910117), and the Army Research Institute (Grant Number: W911NF-22-1-0226).}
}

\begin{document}
%
\maketitle
\begin{abstract}
Understanding why certain individuals work well (or poorly) together as a team is a key research focus in the psychological and behavioral sciences and a fundamental problem for team-based organizations. 
Nevertheless, we have a limited ability to predict the social and work-related dynamics that will emerge from a given combination of team members. 
In this work, we model vocal turn-taking behavior within conversations as a parametric stochastic process on a network composed of the team members.
More precisely, we model the dynamic of exchanging the `speaker token' among team members as a random walk in a graph that is driven by both individual level features and the conversation history.
We fit our model to conversational turn-taking data extracted from audio recordings of multinational student teams during undergraduate engineering design internships. 
Through this real-world data we validate the explanatory power of our model and we unveil statistically significant differences in speaking behaviors between team members of different nationalities.
\end{abstract}
%
%

\section{Introduction}

Data consisting of entities in interconnected systems are ubiquitous in multiple fields.
Thus, network structures are commonly used across many disciplines for representation and analysis of complex information~\cite{brugere2018network}, from neuroscience~\cite{Medaglia1373} to wireless communications~\cite{chowdhury2021}.
In this work, we represent interactions within teams as small in-person social networks.

Computational models can be an effective way to study the social dynamics that emerge from individuals interacting within groups~\cite{McGrath2000}. 
In particular, a variety of modeling approaches have been used to try to replicate natural turn-taking behaviors observed in conversation~\cite{Parker1988, Basu, Stasser1991, Padilha2002}. 
Many of these models have been successful in replicating realistic patterns of conversational turn-taking. 
However, the ability to understand the driving mechanisms underlying these patterns and generalize to novel team compositions is lacking. 
As a step towards this goal, we develop a stochastic model of conversations that can be used to explore how individual differences impact the emergence of turn-taking patterns within teams. 
More precisely, we propose a parametric model that captures the individuals' innate tendency to speak as well as the effect that having spoken recently has on speaking again.
At every point in time, the next speaker is drawn from a probability distribution determined by the history of speakers and the aforementioned parameters.
The model replicates the majority of conversational turn-taking patterns observed in our real-world data, and our results highlight the important role the memory function plays in replicating these patterns. 
Furthermore, our results indicate that differences in team member nationality can play a strong role in shaping communication patterns within multinational teams.

\vspace{2mm}
\noindent
{\bf Contributions.}
The contributions of our work are twofold:\\
i) We propose a simple parametric stochastic process that can capture complex behaviors observed in real data.\\
ii) We present a novel dataset of conversational turn-taking in undergraduate teams and we apply our model to reveal significant differences in speaking behavior between student nationalities. 

\section{Conversation Model}

Inspired by a model by Stasser and Taylor \cite{Stasser1991}, our model incorporates two key notions: i)~the relative likelihood $\pi_i$ that team member $i$ speaks on a given turn independent of their speaking history and
ii)~the effect $m_i$ that an individual’s current speaking turn has on their likelihood of speaking on subsequent turns. 
We consider the inherent likelihood of speaking $\pi_i$ of each member $i$ as independent of the history of exchanges and, thus, a constant throughout the conversation.
In contrast, we encode dependencies within a conversation through the (turn-dependent) memory function $m_i(t)$.
More precisely, for a given member $i$ and a turn $t$, the memory value is given by
\begin{equation}\label{eq:memory}
    m_i(t) = d_i \, e^{{-0.5 \, (t - t_i^{\mathrm{last}})}},
\end{equation}
where $d_i$ is a learnable parameter that controls the scale of the memory effect for each individual and $t_i^{\mathrm{last}}$ denotes the last turn on which member $i$ spoke.
The negative exponential form in~\eqref{eq:memory} reveals that the memory value asymptotically decreases to $0$ as $(t - t_i^{\mathrm{last}})$ increases, i.e., as more turns occur since the last time that member $i$ spoke.
This encodes the natural assumption that whether or not an individual spoke many turns ago is inconsequential to their likelihood of speaking next.
The memory function in~\eqref{eq:memory} is combined with the innate speaking tendencies $\pi_i$ to compute the likelihood $\ell_i(t)$ that member $i$ speaks at turn $t$ as follows
\begin{equation}\label{eq:speaking_likelihood}
\ell_i(t) = 
\begin{cases}
0, \qquad  \qquad \,\,\,\,\, \text{if} \,\,\, t_i^{\mathrm{last}} = t-1,\\
\pi_i + m_i(t), \quad \text{otherwise}.
\end{cases}
\end{equation}
Speakers are not allowed to speak on two consecutive turns since these would simply be considered part of the same turn.
This is enforced in~\eqref{eq:speaking_likelihood} by setting the likelihood to zero for the member that has just spoken.
Lastly, denoting by $N$ the total number of team members, the likelihoods $\ell_i(t)$ are normalized to sum up to $1$ so that they define bona fide probabilities $p_i(t)$ as follows
\begin{equation}\label{eq:speaking_probabilities}
    p_i(t) = \frac{\ell_i(t)}{\sum_{j=1}^{N} \ell_j(t)}.
\end{equation}
The speaker at turn $t$ is then drawn from this probability distribution across team members.

In summary, the conversational behavior of each individual $i$ within our model is given by two parameters $(\pi_i, d_i)$.
Given these parameters for every team member, the model provides a well-defined stochastic process to generate conversations by the team.
More precisely, to determine the speaker at turn $t$, we first compute the memory values of each member following~\eqref{eq:memory}, we then compute likelihoods and transform those into probabilities following~\eqref{eq:speaking_likelihood} and~\eqref{eq:speaking_probabilities}, respectively, and we finally draw the next speaker from that probability distribution.

Our main departure from Stasser and Taylor \cite{Stasser1991} is that our model is based on individual-level parameters whereas theirs is based on team-level parameters. Specifically, Stasser and Taylor’s \cite{Stasser1991} model depends on a single parameter $r$ that determines the inherent speaking probability of every individual (what we denote by $\pi_i$) as well as a single parameter $d$ that determines the scale of the memory function for every team member. This fundamental difference is a key enabler for our study of how individual traits relate to each team member’s conversational behavior since, given observed conversations of a team, our model enables the estimation of the parameters $(\pi_i, d_i)$ for every team member. 

Given observed turn-taking data, we can fit our model by selecting the parameters $(\pi_i, d_i)$ for every team member that maximize the probability of generating the observed data.
More precisely, if we denote by $\mathcal{H}_{t-1}$ the history of turn-taking up to turn $t-1$ in the observed data for a given team, and by $h_t$ the speaker at turn $t$, we can compute the probability that our model selects that true speaker $h_t$ [cf.~\eqref{eq:speaking_probabilities}].
Following the notation in~\eqref{eq:speaking_probabilities}, we denote this probability by $p_{h_t}(t \,| \,\mathcal{H}_{t-1}, \{(\pi_i, d_i)\}_{i=1}^n)$, i.e., the probability of selecting the true speaker $h_t$ at turn $t$ but where we have now made explicit that this value depends on the past history $\mathcal{H}_{t-1}$ and the parameters $(\pi_i, d_i)$ for each of the $n$ members in the team.
With this notation in place, the log-likelihood of observing the true history of $T$ turns is given by
\begin{equation}\label{eqn:loglikelihood}
\mathcal{L}(\mathcal{H}_T \,| \, \{(\pi_i, d_i)\}_{i=1}^n) = \sum_{t=1}^T \log p_{h_t}(t \,| \,\mathcal{H}_{t-1}, \{(\pi_i, d_i)\}_{i=1}^n).
\end{equation}
We fit our model by finding the parameters $\{(\pi_i, d_i)\}_{i=1}^n$ that maximize~\eqref{eqn:loglikelihood}.
We also fit a \emph{reduced model} that does not contain the memory parameters $\{d_i\}_{i=1}^n$ or, equivalently, where $d_i=0$ for all $i$.
We did this to determine the minimal viable model that can explain our observed data.

\section{Dataset}
\label{sec:dataset}

In 2016 and 2017, we collected data on team interactions in student engineering design teams during 7-week internships at a private university in the southern United States. 
The first week of the internship consisted of a condensed course on the engineering design process, which helped to ensure all participants had a similar baseline level of knowledge. 
During the remaining six weeks, team members worked together to plan and execute their project which sought to meet a real-world need. 
We collected data from 7 multi-national teams with team members from the United States (n = 13), Malawi (n = 7), and Brazil (n = 4), with equal numbers of female and male participants. 
After consenting to the study, participants completed a self-report survey of their personality traits, attitudes, and demographic information. 
From these data we extracted five features for each individual that we hypothesized could relate to individual differences in speaking patterns, namely, extraversion, agreeableness, conscientiousness, sex (male, female), and nationality (American, Non-American).
Our dataset includes multiple meetings from throughout the internships for all teams. Audio streams were processed by annotating the start and end times of speaking turns by each team member during the meetings. Overall, we extracted a mean (SD) of 1941 (1416.5) speaking turns per team.

\section{Numerical experiments}

 To assess the predictive power of the full and reduced (i.e., without the memory component) models, we perform the following three classes of experiments. 


\vspace{1mm}
\noindent
{\bf Predicting the next speaker.}
For each team, we split their turn-taking history into a training and a testing set. 
The training data contains the first 80 percent of the total turns whereas the testing data contains the remaining turns. 
As previously explained, we compute the maximum likelihood estimates of the model parameters but this time based only on maximizing the probability of observing the history of the training dataset. 
We then compute the log-likelihood of observing the history of the testing dataset, as in (4), for both the full and reduced fitted models. 
The larger (less negative) the attained value, the better predictive power of the corresponding model.

Overall, the full simulation model (i.e., with memory parameter) predicts the observed data better than the reduced simulation model. 
Table~\ref{tab:table1} shows the log-likelihoods attained for the testing dataset (last 20 percent of speaking turns) for each team and simulation model. 
The full simulation model consistently yields larger (less negative) log-likelihoods, indicating a better predictive performance.

\begin{table}[t]
\centering
    \caption{Log-likelihood attained by both models}
    \label{tab:table1}
    \begin{tabular}{c c c} 

      Team & No Memory & Memory \\
      \hline
      Team 1 & -159.1586 & -154.4533 \\
      Team 2 & -45.3267 & -43.2123 \\
      Team 3 & -196.5418 & -188.6291 \\
      Team 4 & -278.8524 & -255.2545 \\
      Team 5 & -909.8953 & -617.3478 \\
      Team 6 & -623.9848 & -589.0674 \\
      Team 7 & -105.8199 & -104.0864 \\
    \end{tabular}
    \vspace{-0.3cm}
\end{table}

\vspace{1mm}
\noindent
{\bf Reproducing relevant conversation patterns.}

We test the fit of both the full and reduced simulation models by comparing three measures between our observed and simulated datasets: 
1) the proportion of time in which each team member spoke, 
2) the proportion of a given speaker’s speaking turns following an ABA format in which there was one turn by a different speaker between a given speaker’s sequential turns (reflecting the proportion of turns that were part of dyadic exchanges), and 
3) the proportion of turns that were part of long dyadic exchanges (4 or more consecutive speaking turns (e.g. ABAB) between two team members). 
We calculate these measures for each of 10,000 replications of our simulation models and compare them to the values found in our observed data from each team.
For each of our three measures, we find the proportion of model replications in which the observed value in the real data fell within the 95 percent confidence interval for values produced by each simulation model. 
For each of the three speaking patterns of interest, we use chi-squared tests to compare the number of individuals or dyads across teams whose behaviors are not significantly different from those displayed by the full and reduced simulation models.

\begin{figure*}[ht]
		\centering
  \includegraphics[width=0.95\textwidth]{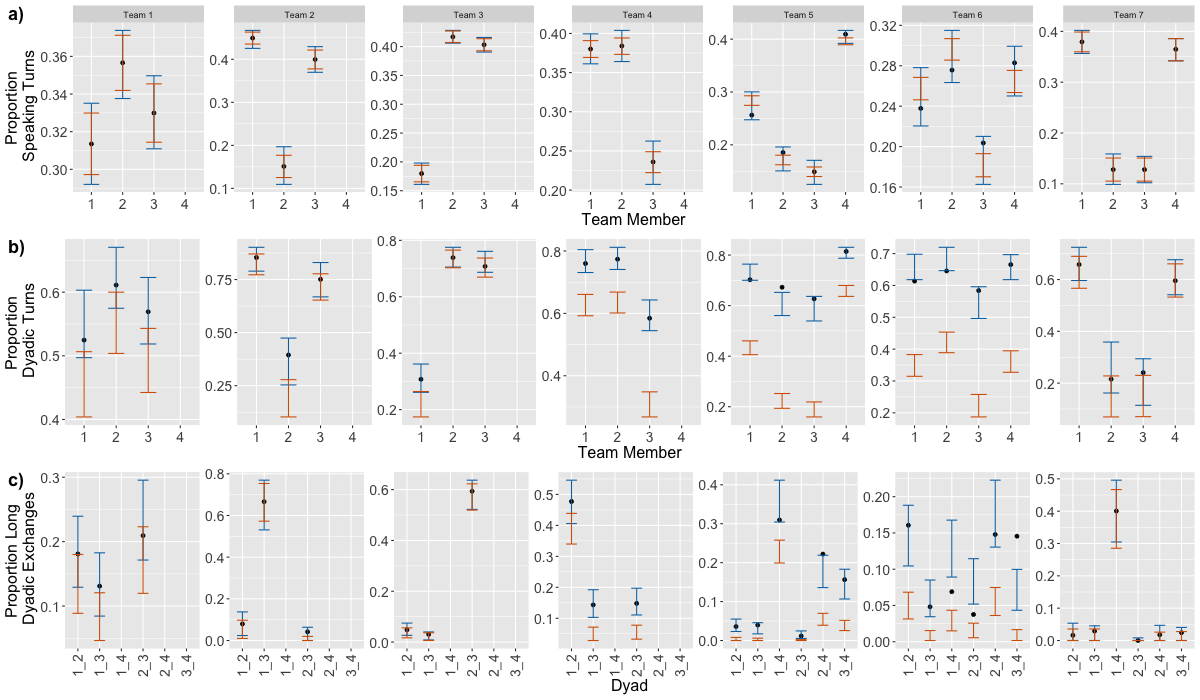}
	\caption{\small{Black points represent a) observed proportion of speaking turns by each team member, b) observed proportion of speaking turns with one turn in between (e.g. ABA) for each team member, c) observed proportion of speaking turns that were part of consecutive dyadic exchanges of length 4 turns or greater. Error bars represent the 95 percent confidence intervals for the proportions estimated by the reduced simulation model (i.e., without memory parameter) (red) and full simulation model (blue). Y-axis scale varies by team to improve visibility}}
	\label{F:prop_speaking_time}
\end{figure*}

The full simulation model matches the patterns displayed by significantly more individuals and dyads across teams than the reduced simulation model. 
The full simulation model correctly estimates the proportion of speaking turns spoken by each team member for $100 \%$ (24/24) of team members whereas the reduced simulation model correctly estimates the proportion for $70.8 \%$  (17/24) of team members ($\chi^2$ = 6.0, p = 0.014; Figure~\ref{F:prop_speaking_time}(a)). 
Moreover, the full simulation model correctly estimates the proportion of each team member’s speaking turns with an ABA format for $87.5 \%$  (21/24) of team members, but the reduced simulation model correctly estimates the proportion for $29.2 \%$ (7/24) of team members, with the tendency to underestimate the proportion of turns ($\chi^2$ = 14.5, p = 0.00014; Figure~\ref{F:prop_speaking_time}(b)). 
Finally, the full simulation model correctly estimates the proportion of speaking turns that were part of dyadic exchanges of length 4 turns or greater for $86.7 \%$ (26/30) of team member dyads, and the reduced simulation model correctly estimates the proportion for $36.7 \%$  (11/30) of team member dyads, with the tendency to underestimate the proportion of turns ($\chi^2$ = 13.8, p = 0.00020; Figure~\ref{F:prop_speaking_time}(c)).

\vspace{1mm}
\noindent
{\bf Relating individual traits and speaking behavior.}
To gain insight into the relative importance of different individual traits in understanding speaking behaviors, we use an information-theoretic approach~\cite{Burnham2002} to determine which trait(s) best explain between-individual variation in model parameters $\pi_i$ (baseline likelihood of speaking) and $d_i$ (likelihood of speaking again after speaking recently). 
Using the MuMIn function~\cite{Barton2013} in R, we examine which linear model (i.e., a null model and 5 uni-variate models consisting of each of our individual-level predictor variables; see Section~\ref{sec:dataset}) best explains variation in each parameter value across team members. We group-mean center our three continuous variables (extraversion, agreeableness, conscientiousness) to reflect the relative values of these personality traits among team members.
We limit the number of variables per linear model to one to avoid overfitting.
We rank our linear models according to the Akaike information criterion adjusted for small sample sizes (AICc)~\cite{Burnham2002}. 
We consider top-performing linear models to be the best performing model (i.e. the model with the lowest AICc value) in our model set and any model less than 2 AICc points greater than the best performing linear model~\cite{Burnham2002}.
We examine the correlation between individual traits and simulation model parameters for all top-performing linear models to determine how these traits shape speaking behaviors. 

Tables~\ref{tab:table2} and ~\ref{tab:table3} display the results of our model selection analysis that compares the relative ability of each of our five univariate models to explain between-individual variation in $\pi_i$ and $d_i$. The tables display the AICc values for each model as well as the $\Delta$AICc values (relative to the best model) and corresponding model weights. Model weights reflect the relative support for a given linear model compared to the other candidate models, with 1 indicating full support. 

\begin{table}[h!]
     \caption{\label{tab:table2} Model Selection for Baseline Likelihood of Speaking $\pi_i$}
     \begin{center}
    \begin{tabular}{l|l|c|c|c|c|c|l|l|l} 

      model & df & AICc & $\Delta$ & wt  \\
      \hline
      Nationality & 3 & -19.6 & 0.0 & 0.94\\
      Null & 2 & -12.4 & 7.2 & 0.025\\
      Agreeableness & 3 & -11.4 & 8.2 & 0.015\\
      Sex & 3 & -10.3 & 9.3 & 0.010\\
      Extraversion & 3 & -10.0 & 9.6 & 0.010\\
      Conscientiousness & 3 & -9.9 & 9.7 & 0.010\\
    \end{tabular}
    \end{center}
    \vspace{-0.3cm}
\end{table}

\begin{table}[h!]
    \caption{Model Selection for Shape of Memory Function $d_i$}
    \label{tab:table3}
    \begin{center}
    \begin{tabular}{l|l|c|c|c|c|c|l|l|l} 

      model & df & AICc & $\Delta$ & wt  \\
      \hline
      Null & 2 & 105.2 & 0.0 & 0.42\\
      Extraversion & 3 & 107.5 & 2.4 & 0.13\\
      Nationality & 3 & 107.7 & 2.5 & 0.12\\
      Conscientiousness & 3 & 107.8 & 2.6 & 0.11\\
      Sex & 3 & 107.8 & 2.6 & 0.11\\
      Agreeableness & 3 & 107.8 & 2.6 & 0.11\\
    \end{tabular}
    \end{center}
\end{table}

\begin{figure}
  \begin{center}
    \includegraphics[width=0.3\textwidth]{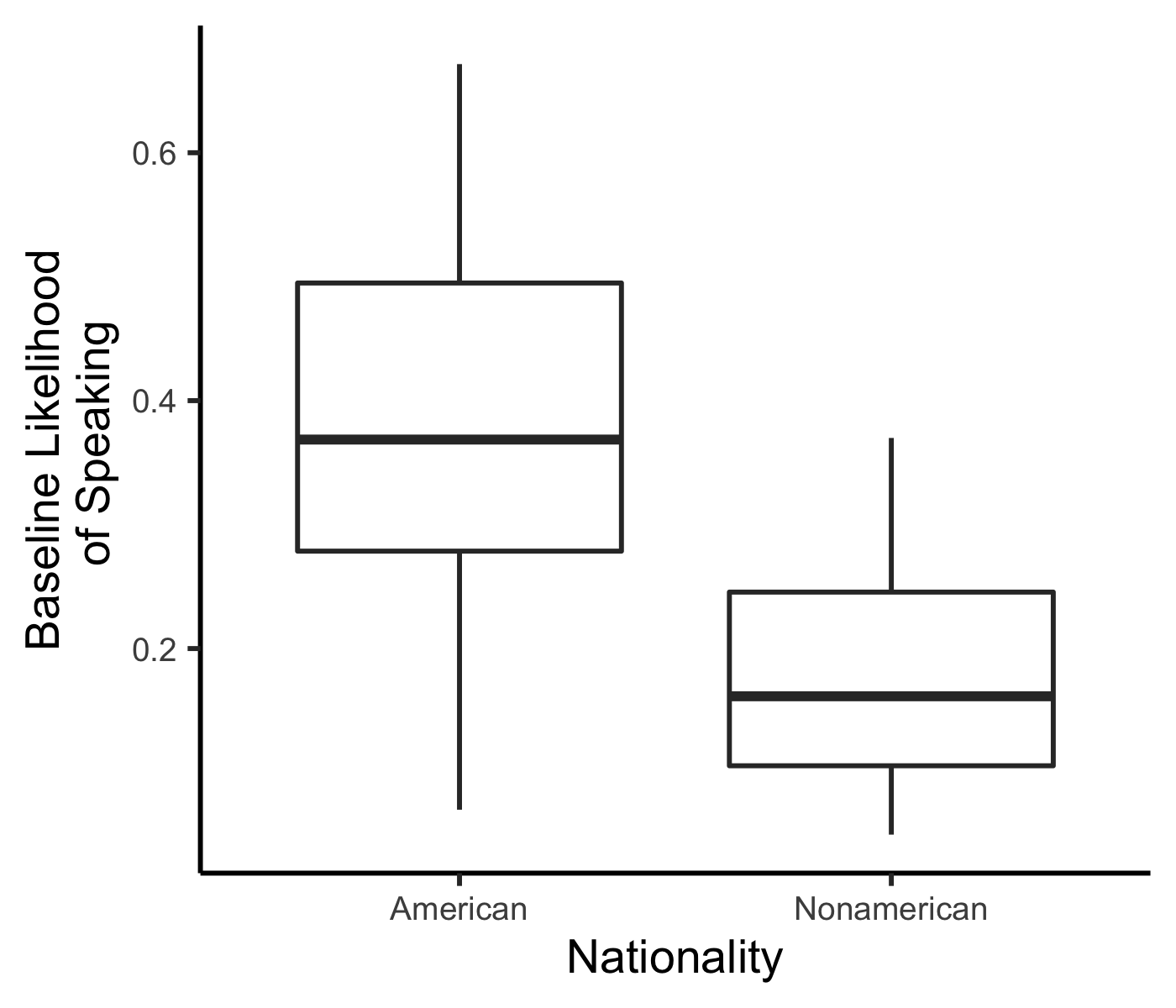}
  \end{center}
  \vspace{-4mm}
  \caption{\small Boxplot of baseline likelihood of speaking (parameter $\pi_i$) by nationality across all teams.}
  \label{fig:nationality}
\end{figure}

The linear model that best explains between-individual differences in $\pi_i$, the parameter reflecting the baseline likelihood of initiating a speaking turn, has nationality as the predictor variable. 
This linear model has a cumulative model weight of $93.5 \%$. The second best linear model is the null model, which has a $\Delta$AICc value 7.2 higher than the best model (Table~\ref{tab:table2}). 
Since this $\Delta$AICc value is greater than our criterion of $\Delta$AICc = 2 \cite{Burnham2002}, we only consider the linear model with nationality as a predictor variable as a top-performing model within our model set.
Overall, this model is supported 37.6 times more strongly (evidence ratio = $w_i$/$w_j$ = 0.94/0.025 = 37.6) than the null model. 
When we analyze our top model, we find that Americans have significantly higher likelihoods of initiating speaking turns than non-Americans ($\beta$ = 0.20, p $<$ 0.01, Figure~\ref{fig:nationality}).

The linear model that best explains between-individual differences in $d_i$, the change in likelihood of speaking after having just spoken, is the null model. 
This linear model has a cumulative model weight of $41.6 \%$. The second best linear model has extraversion as a predictor variable and a $\Delta$AICc value of 2.4 (Table~\ref{tab:table3}).
Thus, only the null model is considered a top-performing linear model within our model set, indicating that none of the predictor variables we considered explain between-individual variation in $d_i$.
Overall, the null model is supported approximately 3.2 times more strongly (evidence ratio = $w_i$/$w_j$ = 0.42/0.13 = 3.2) than the model with extraversion as a predictor.



\section{Discussion}

The presence of the memory parameter is important in simulating the patterns of vocal turn-taking we observed in our study. 
Compared to the reduced simulation model with no memory parameter, the full simulation model more accurately predicts future speaking turns and better captures individual and dyadic speaking patterns.
This result supports the findings by Stasser and Taylor~\cite{Stasser1991} and Parker~\cite{Parker1988} that an individual's current likelihood of speaking is impacted by their recent speaking behaviors.
Nevertheless, our results differ from those of Stasser and Taylor in that we find different parameter values controlling memory function shape for different individuals.

Our study finds evidence for consistent between-individual differences in speaking behaviors supporting previous findings that individual traits can correlate with communication behaviors~\cite{Li2019, Mast2002, Kimble1988}. 
Our finding that non-Americans initiate speaking turns less frequently than Americans is consistent with a recent study by Li et al.~\cite{Li2019} which found that Chinese team members, who tended to be less proficient in English, initiated fewer speaking turns than the American team members.
Although we did not measure English language proficiency in our study, our finding could be related to language proficiency since the non-American students in our study were non-native English speakers.
Another reason why non-Americans may not have initiated speaking turns as frequently could be that they had a perceived lower status than American team members. 
Social status may be awarded to the ethnic subgroup with the greatest numerical majority~\cite{Bellmore2007}.
Since both the Brazilian and Malawian students were completing the internship at an American university and were outnumbered by American students, they may have demonstrated lower status behaviors like speaking up less frequently~\cite{Mast2002}.

Although nationality best explained differences in baseline frequency of speaking turn initiation, none of our predictor variables explained variation in memory function shape. 
Future studies are needed to determine whether other traits may explain the observed variation in this speaking tendency.  
Nevertheless, since Americans were more likely to initiate speaking turns, the broad tendency to speak again after having recently spoken further enhanced individual differences in speaking frequency across team members.  
Overall, these results help expand knowledge of the impact cultural diversity can have on team processes~\cite{Stahl2010}.

A limitation of our study was that we only had data on a relatively small number of teams and team members. 
This lack of power prevented us from exploring more complex relationships between individual traits and their impacts on speaking behaviors.
For example, Neubert and Tagger~\cite{Neubert2004} found that gender moderated the relationship between individual traits and leadership, with certain traits being more important for leadership in males than females and vice versa.
Since we tested each of our predictor variables on its own, the strong effect of nationality may have overpowered more subtle or complicated effects of other variables, such as personality and gender. 
This could be a reason why individual traits like extraversion, which can be strongly correlated with communication tendencies~\cite{McLean2006, Leung2001, Macht2014}, did not correspond to individual differences in speaking behaviors in our study.
Extending our study to more teams would enable a greater understanding of how multiple traits may interact to impact speaking behaviors.

\section{Conclusion and future work}

Our study develops a model of conversational turn-taking that can provide a mechanistic understanding of how patterns of communication emerge within teams and can be used to investigate the relationship between team member traits and specific speaking behaviors. 
Future extensions of our model could integrate more fine-grained speaking behaviors such as the timing between turns and turn overlap, which may enable the study of more complex or subtle turn-taking dynamics. For example, individuals higher in dominance have been found to interrupt more often, which can have a suppressive effect on the speaking behaviors of others~\cite{Rogers1970}.
Ultimately, through extensions of our modeling approach, it could be possible to predict the conversational interactions among team members based on their trait composition alone. This ability could enable the anticipation of undesirable team outcomes (e.g., development of subgroups) so that interventions could be applied ahead of time. 
Similarly, for established teams, it could also be possible to predict the effects team composition changes may have on communication patterns, thus providing guidelines for restaffing or retraining team members.
Such predictive models would represent a significant advancement in teams research, enabling a more mechanistic understanding of the connection between team composition and team processes~\cite{Bell2018a, OBryan2020}.

\bibliographystyle{IEEEtran}
\bibliography{myIEEEabrv,biblio}

\end{document}